\documentclass[twocolumn,preprintnumbers,pra,amsmath,amssymb,superscriptaddress,showpacs]{revtex4}
\usepackage{epsfig}
\usepackage{graphicx}

\begin{document}

\title{Tight bounds on the concurrence of quantum superpositions}

\author{J. Niset}
\affiliation{Quantum Information and Communication, Ecole Polytechnique, CP 165, Universit\'e Libre de
Bruxelles, 1050 Brussels, Belgium}

\author{N. J. Cerf}
\affiliation{Quantum Information and Communication, Ecole Polytechnique, CP 165, Universit\'e Libre de
Bruxelles, 1050 Brussels, Belgium}

\begin{abstract}
The entanglement content of superpositions of quantum states is investigated
based on a measure called {\it concurrence}. 
Given a bipartite pure state in arbitrary dimension written as
the quantum superposition of two other such states, we find simple inequalities 
relating the concurrence of the state to that of its components. 
We derive an exact expression for the concurrence when the component
states are biorthogonal, and provide elegant upper and lower bounds
in all other cases. For quantum bits, our upper bound is tighter than
the previously derived bound in [Phys. Rev. Lett. 97, 100502 (2006).]  
\end{abstract}

\pacs{03.67.-a, 03.67.Mn}

\maketitle
\section{Introduction}
Consider a quantum state $\vert\Psi\rangle$ of two parties, Alice and Bob, written as the superposition of two terms 
\begin{equation}\label{sup}
\vert\Psi\rangle=\alpha\vert\phi\rangle+\beta\vert\varphi\rangle
\end{equation}
with $\vert\alpha\vert^2+\vert\beta\vert^2=1$. We are interested in the relation between the entanglement of $\vert \Psi \rangle$ and that of $\vert\phi\rangle$ and $\vert\varphi\rangle$. This problem was recently addressed by Linden, Popescu, and Smolin in Ref.~\cite{Linden}, where an upper bound on the entanglement of $\vert\Psi\rangle$ 
was derived using the {\it entanglement of formation} $E_\mathrm{f}(\Psi)$ 
as a measure of entanglement, i.e., the von Neumann entropy 
of the reduced state 
of either party. When Alice and Bob hold 2-dimensional quantum systems
(qubits), the entanglement of formation $E_\mathrm{f}(\Psi)$ completely
characterizes the entanglement of the pure state $\vert \Psi \rangle$. However, for higher-dimensional systems (and in the case of a finite number of copies), the measure of entanglement is not unique, and $E_\mathrm{f}(\Psi)$ alone is not sufficient to completely describe the entanglement of a bipartite pure state. With a pair of $3$-dimensional systems, for example, one can find states that have the same value of $E_\mathrm{f}(\Psi)$ but different Schmidt numbers, 
hence cannot be transformed to each other by Local Operations and Classical Communications \cite{Horod1}. Although the two states have the same amount of entanglement, as measured by the number of singlet states needed (asymptotically) to prepare them, the nature of this entanglement is quite different. Indeed, one can argue that for a pair of $d$-dimensional systems, the structure of entanglement depends on $d-1$ independent Schmidt coefficients, hence a complete description would require $d-1$ independent measures of entanglement \cite{Vidal}. 

In this paper, we pursue the study of the entanglement of quantum 
superpositions based on another measure of entanglement
than $E_\mathrm{f}(\Psi)$.
This study is thus particularly interesting in dimensions higher than 2. 
In Ref.~\cite{Linden}, an exact solution to this problem was derived in the
special case where the two component states are biorthogonal. 
In this scenario, one may ask whether such an exact solution can also be
obtained with other measures of entanglement. In addition, in the  
cases where the component states are orthogonal (but not biorthogonal) 
or even arbitrary, can we derive new upper bounds, 
as was done for $E_\mathrm{f}(\Psi)$ in \cite{Linden}, 
but also lower bounds? These questions 
are of great importance since a good characterization of the entanglement of a superposition clearly requires both the knowledge of upper and lower bounds. 
Furthermore, one would like to know whether one can extract a common structure for the bounds based on different measures of entanglement as this would highlight the mechanism underlying the entanglement of superpositions, hence improve our understanding of entanglement itself. 

In what follows, we shall address these various issues 
based on another widely used measure of entanglement, the so-called 
{\it concurrence}~\cite{Wootters}. For 2-level systems, 
which are studied in Section II, this quantity is in one-to-one correspondence
with $E_\mathrm{f}(\Psi)$ so that our results can be directly compared to those
of Ref.~\cite{Linden}. We shall see that our derived upper bound is generally
tighter than that of Ref.~\cite{Linden}, while it can also be supplemented
with a lower bound so that we get strong constraints on the
allowed values for the entanglement of superpositions.
In Section III, we shall generalize our analysis to the case of a pair
of $d$-level systems with any $d$, and derive simple upper and lower
bounds for arbitrary states. The special cases
of biorthogonal and orthogonal component states will also be treated,
and an exact solution will again be given in the former case. As emphasized
above, the derived upper and lower bounds yield news constraints 
to the entanglement of superpositions in $d$ dimensions, 
which are complementary to those of Ref.~\cite{Linden}.

Very recently, the problem of finding bounds on the concurrence 
of superpositions has also been addressed in Ref.~\cite{Yu}. 
Since it is based on matrix notations for pure states, this approach gives,
however, quite complicated bounds, which, in addition, assume the prior knowledge of parameters such as the rank of the superposition state 
or the largest eigenvalues of the component states. 
The usefulness of such bounds is then very questionable as it may become
equally simple to compute directly the entanglement of 
the superposition state $\vert\Psi\rangle$ itself.
To be sensible, the bounds should only depend on the entanglement of the
component states $\vert\phi\rangle$ and $\vert\varphi\rangle$ 
as well as on simple parameters such as 
the coefficients $\alpha$ and $\beta$ or the scalar product $\langle\phi\vert\varphi\rangle$ between the component states. 

\section{Superposition of bipartite pure states in dimension two}

Let us first consider the simplest scenario of Alice and Bob both having a qubit as it will help us gain some useful intuition. For qubits, 
the definition of the concurrence makes use of the {\it spin flip}
transformation $S$ acting on a single qubit,
\begin{equation}
\vert\tilde{\psi}\rangle=S(\vert \psi \rangle)=\sigma_y K (\vert \psi \rangle)
=\sigma_y \vert \psi^*\rangle
\end{equation}
where $K$ denotes the complex conjugation, that is,
$\vert\psi^*\rangle$ is the complex conjugate of  $\vert\psi\rangle$ 
when it is expressed in the eigenbasis of $\sigma_z$. Here, $\sigma_i$
denote the Pauli matrices. With this notation, the concurrence reads \cite{Wootters}
\begin{equation}
C(\Psi)=\vert \langle \Psi \vert \tilde{\Psi}\rangle\vert = \vert\langle\Psi\vert\sigma_y \otimes \sigma_y \vert \Psi^*\rangle\vert
\label{concurrence}
\end{equation}
and it corresponds, physically, to the overlap between the state and its
image under a spin-flip of both qubits. Note that the spin-flip of both qubits can also be interpreted as complex conjugation in the so-called "magic" basis introduced in Ref.~\cite{Wootters}. For two-level systems, the concurrence
has the interesting property of being directly related 
to the entanglement of formation through 
\begin{equation}\label{Eof}
E_\mathrm{f}(\Psi)=h\left( \frac{1+\sqrt{1-C^(\Psi)}}{2}\right)
\end{equation}
where $h(x)$ is the binary entropy function \cite{Wootters}. This simple 
formula will make it possible to connect our results 
to those of Ref.~\cite{Linden}.
\par

Introducing relation (\ref{sup}) in definition (\ref{concurrence}), 
we can write the concurrence of the superposition as 
\begin{align}\label{Conc}
\nonumber C(\Psi)&=\big\vert (\alpha^*)^2\langle\phi\vert\sigma_y\otimes\sigma_y\vert\phi^*\rangle + (\beta^*)^2\langle\varphi\vert\sigma_y\otimes\sigma_y\vert\varphi^*\rangle\\
&+(\alpha^*\beta^*)(\langle\phi\vert\sigma_y\otimes\sigma_y\vert\varphi^*\rangle+\langle\varphi\vert\sigma_y\otimes\sigma_y\vert\phi^*\rangle)\big\vert
\end{align}

\subsection{Orthogonal states} 
First we note that when $\vert\phi\rangle$ and $\vert\varphi\rangle$ are biorthogonal, the problem boils down to the trivial solution $C(\Psi)=2|\alpha\beta|$. The easiest interesting scenario is when the two component
states are orthogonal (but not necessarily biorthogonal), that is, $\langle\phi\vert\varphi\rangle=0$. This
condition implies that we can construct an orthonormal basis $\{\vert\phi\rangle,\vert\varphi\rangle,\vert\xi_1\rangle,\vert\xi_2\rangle\}$ such that the coefficients of any normalized state in this basis will sum to one, i.e., $\forall \vert\psi\rangle$,
\begin{equation}
\vert\langle\phi\vert\psi\rangle\vert^2+\vert\langle\varphi\vert\psi\rangle\vert^2+\vert\langle\xi_1\vert\psi\rangle\vert^2+\vert\langle\xi_2\vert\psi\rangle\vert^2=1
\end{equation}
This is true in particular for the two spin-flipped states $\vert\tilde{\phi}\rangle=\sigma_y\otimes\sigma_y\vert\phi^*\rangle$ and $\vert\tilde{\varphi}\rangle=\sigma_y\otimes\sigma_y\vert\varphi^*\rangle$, hence 
\begin{align}
\nonumber\vert\langle\varphi\vert\sigma_y\otimes\sigma_y\vert\phi^*\rangle\vert\leq\sqrt{1-C^2(\phi)}\\
\vert\langle\phi\vert\sigma_y\otimes\sigma_y\vert\varphi^*\rangle\vert\leq\sqrt{1-C^2(\varphi)}
\label{lhs}
\end{align}
In addition, since $\sigma_y K = -K \sigma_y$, one can easily check that both left-hand sides of (\ref{lhs}) are equal. We thus derive the useful relation
\begin{equation}\label{relorth}
\vert\langle\phi\vert\sigma_y\otimes\sigma_y\vert\varphi^*\rangle\vert=\vert \langle\varphi\vert\sigma_y\otimes\sigma_y\vert\phi^*\rangle\vert \leq \sqrt{1-\delta^2}
\end{equation}
where $\delta=\max(C(\phi),C(\varphi))$.\\

\textbf{Proposition 1: Upper bound.} Let Alice and Bob have a qubit, and let $\vert\phi\rangle$ and $\vert\varphi\rangle$ be orthogonal, the concurrence of the superposition $\vert\Psi\rangle=\alpha\vert\phi\rangle+\beta\vert\varphi\rangle$, with $\vert\alpha\vert^2+\vert\beta\vert^2=1$ satisfies
\begin{align}\label{sup_2}
C(\Psi)&\leq \vert\alpha\vert^2 C(\phi) + \vert\beta\vert^2 C(\varphi)+ 2\vert\alpha\beta\vert\sqrt{1-\delta^2}
\end{align}
where $\delta=\max(C(\phi),C(\varphi))$.\\

\textit{Proof.} Successive application of the triangle inequality (TI) $\vert x+y \vert \leq \vert x \vert+\vert y\vert$ to equation (\ref{Conc}), followed by the introduction of relation (\ref{relorth}) together with the definition of the concurrence of $\vert\phi\rangle$ and $\vert\varphi\rangle$ directly leads  to the upper bound (\ref{sup_2}). $\square$\\

\textbf{Proposition 2: Lower bound.} Under the same conditions and with the same definition of $\delta$, the concurrence of the superposition satisfies
\begin{align}\label{inf_2}
C(\Psi)&\geq \big\vert \vert\alpha\vert^2 C(\phi) - \vert\beta\vert^2 C(\varphi)\big\vert- 2\vert\alpha\beta\vert\sqrt{1-\delta^2}
\end{align}

\textit{Proof.} This time, we make use of the inverse triangle inequality (ITI) $\vert x+y \vert \geq \big\vert \vert x \vert-\vert y\vert\big\vert$. First, we apply it to (\ref{Conc}) and separate between the first two and the last two terms of the right-hand side. Next, we remember that the absolute difference of two positive terms is always greater than their difference, i.e. $\big\vert \vert x \vert-\vert y\vert\big\vert \geq \vert x \vert-\vert y\vert$. We can then express our bound by applying ITI again to the first of these positive terms, and TI to the second. Finally, we inject relation (\ref{relorth}) and the definition of the concurrence of $\vert\phi\rangle$ and $\vert\varphi\rangle$ into this last expression to obtain the lower bound (\ref{inf_2}). $\square$\\\

\begin{figure}[t]
\centering
\includegraphics[scale=0.28]{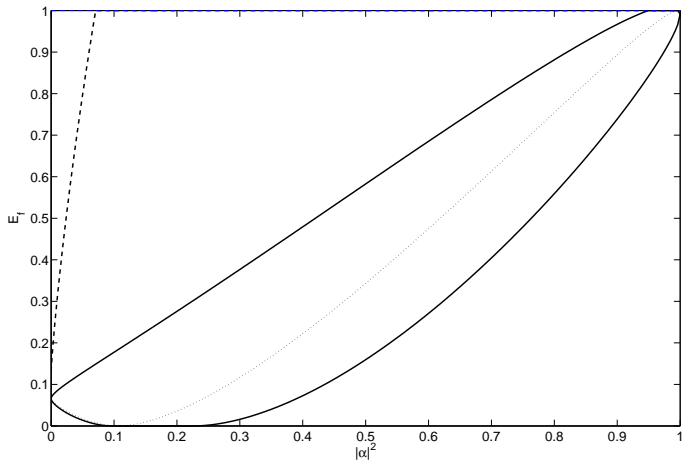}
\caption{Entanglement of formation of $\vert\Psi\rangle=\alpha\vert\phi\rangle+\beta\vert\varphi\rangle$,
	with $\vert\phi\rangle$ and $\vert\varphi\rangle$ being two random orthogonal states defined in the text.
	The dotted line is the exact value of $E_f(\Psi)$, while the two solid lines correspond to the upper and lower bounds derived from relation (\ref{sup_2}) and (\ref{inf_2}). The dashed line is the upper bound of Ref.~\cite{Linden}.}
\label{fig:Eof}
\end{figure}

\indent Let us comment on our results until now.
First, this approach based on the concurrence is fruitful in the sense
that, for two-dimensional systems, it makes it possible to derive both an upper
and a lower bound on the entanglement of a superposition. 
Second, these bounds can be saturated. Consider, for example,
the states 
\begin{align}
\vert\Psi_1\rangle&=\alpha_1\vert\Phi^+\rangle + \beta_1\vert 01 \rangle 
\nonumber \\
\vert\Psi_2\rangle&=\alpha_2\vert\Phi^+\rangle + \beta_2\vert\Phi^-\rangle
\end{align}
with $\alpha_i$ and $\beta_i$ real, and $\vert\Phi^\pm\rangle=1/\sqrt{2}(\vert 00 \rangle \pm \vert 11 \rangle )$. One can easily check that $C(\Psi_1)$  exactly saturates the upper bound (\ref{sup_2}), while $C(\Psi_2)$ exactly saturates the lower bound (\ref{inf_2}). Finally, when translated to a bound
on the entanglement of formation via relation (\ref{Eof}), our upper bound (\ref{sup_2}) typically gives stronger constraints than those derived 
in Ref.~\cite{Linden}. This is illustrated in Fig.~\ref{fig:Eof} for
two orthogonal randomly generated states $\vert\phi\rangle=-0.264\vert 00 \rangle+0.528\vert 01 \rangle+0.487\vert 10 \rangle-0.643\vert 11 \rangle$ and $\vert\varphi\rangle=-0.034\vert 00 \rangle+0.675\vert 01 \rangle-0.734\vert 10 \rangle+0.010\vert 11 \rangle$.

\subsection{Arbitrary states} 

When the two component states are not orthogonal, the superposition (\ref{sup}) is not normalized. We can nevertheless derive bounds on the concurrence of the normalized version of this superposition, i.e., $\vert \Psi'\rangle=\vert \Psi\rangle/\Vert\Psi\Vert$, using the same method as before. Note that relation (\ref{relorth}) does not hold anymore. 
However, we can always introduce an orthonormal basis, say $\{\vert\phi\rangle,\vert\xi_1\rangle,\vert\xi_2\rangle,\vert\xi_3\rangle\}$, such that $\langle\varphi\vert\xi_2\rangle=\langle\varphi\vert\xi_3\rangle=0$. Again, the coefficients of any normalized state expressed in this basis will sum to one. After some straightforward calculations, we find that the resulting expressions for the two spin-flipped states  $\vert\tilde{\phi}\rangle=\sigma_y\otimes\sigma_y\vert\phi^*\rangle$ and $\vert\tilde{\varphi}\rangle=\sigma_y\otimes\sigma_y\vert\varphi^*\rangle$
are
\begin{align}
\nonumber\vert\langle\varphi\vert\sigma_y\otimes\sigma_y\vert\phi^*\rangle\vert\leq\sqrt{1-(C(\phi)-\vert\langle\phi\vert\varphi\rangle\vert)^2}\\
\vert\langle\phi\vert\sigma_y\otimes\sigma_y\vert\varphi^*\rangle\vert\leq\sqrt{1-(C(\varphi)-\vert\langle\phi\vert\varphi\rangle\vert)^2}
\end{align}
Noting again that the two left-hand side terms are equal, we have
\begin{align}\label{relarbitrary}
\vert\langle\phi\vert\sigma_y\otimes\sigma_y\vert\varphi^*\rangle\vert=\vert \langle\varphi\vert\sigma_y\otimes\sigma_y\vert\phi^*\rangle\vert \leq \sqrt{1-\delta^2}
\end{align}
with $\delta=\max\Big(\big| C(\phi)-\vert\langle\phi\vert\varphi\rangle\vert\big|,\big| C(\varphi)-\vert\langle\phi\vert\varphi\rangle\vert\big|\Big)$.\\

\indent\textbf{Proposition 3: Arbitrary states.} When $\vert\phi\rangle$ and $\vert\varphi\rangle$ are arbitrary normalized states, the concurrence of the normalized version of the superposition $\vert\Psi\rangle=\alpha\vert\phi\rangle+\beta\vert\varphi\rangle$ with $\vert\alpha\vert^2+\vert\beta\vert^2=1$ obeys the upper and lower bounds
\begin{align}
\nonumber \Vert\Psi\Vert^2 C(\Psi')&\leq \vert\alpha\vert^2 C(\phi) + \vert\beta\vert^2 C(\varphi) + 2\vert\alpha\beta\vert\sqrt{1-\delta^2}\\
\Vert\Psi\Vert^2 C(\Psi')&\geq \big\vert \vert\alpha\vert^2 C(\phi) - \vert\beta\vert^2 C(\varphi)\big\vert - 2\vert\alpha\beta\vert\sqrt{1-\delta^2}
\end{align}
where $\delta=\max\Big(\big| C(\phi)-\vert\langle\phi\vert\varphi\rangle\vert\big|,\big| C(\varphi)-\vert\langle\phi\vert\varphi\rangle\vert\big|\Big)$.\\

\textit{Proof.} We first realize that $C(\Psi')=\vert\langle\Psi'\vert\sigma_y\otimes\sigma_y\vert\Psi'\rangle\vert=\vert\langle\Psi\vert\sigma_y\otimes\sigma_y\vert\Psi\rangle\vert/\Vert\Psi\Vert^2$, then proceed exactly as before using (\ref{relarbitrary}) instead of (\ref{relorth}). $\square$\\

\section{Systems of arbitrary dimension}

Although the concurrence was initially introduced to measure the entanglement of a pair of qubits, it has since been generalized to pairs of quantum systems of arbitrary dimension $d$. The first generalization, based on the notion of conjugations, is due to Uhlmann \cite{Uhlmann}. However, this approach was later shown to lead to a definition of concurrence that cannot serve as a basis for a general measure of entanglement \cite{Rungta}.  We will thus use in what
follows the definition of the generalized concurrence introduced by Rungta \textit{et al.} \cite{Rungta}, sometimes known as {\it I-concurrence}. This definition makes use of a superoperator called the {\it Universal Inverter},
which naturally extends the idea of spin flip to $d$-dimensional 
quantum systems or qudits (the extension of the spin flip to more 
than 2 dimensions had also been studied in the context of the reduction criterion for separability in Refs.~\cite{Cerf,Horod}). 
On a qudit state $\rho$, the action of the Universal Inverter is given by
\begin{equation}
S_d(\rho)=\nu_d(I-\rho)
\end{equation}
where $\nu_d$ is a positive constant. Thus, the Universal Inverter maps a pure qudit state $\vert\Psi\rangle$ onto a multiple of the maximally mixed state in the subspace orthogonal to $\vert\Psi\rangle$. As argued in \cite{Rungta}, in order to have a meaningful definition of the concurrence, consistent with the expression of the concurrence for qubits, one should choose the scaling factor $\nu_d$ equal to one. However, $\nu_d$ is sometimes chosen equal to $1/(d-1)$ \cite{Albeverio}, and we will briefly discuss this possibility at the end of this section. With $\nu_d=1$, as we assume in what follows, the Universal Inverter is a trace-increasing positive superoperator, which preserves Hermiticity. The corresponding generalized concurrence for a bipartite pure state $\vert\Psi\rangle$ of $d$-dimensional quantum systems is given by
\begin{equation}\label{conc}
C(\Psi)=\sqrt{\langle\Psi\vert S_d\otimes S_d (\vert\Psi\rangle\langle\Psi\vert)\vert\Psi\rangle}
\end{equation}

Let us now briefly state some useful properties of the superoperator $\Lambda_d=S_d\otimes S_d $ with $\nu_d=1$, as defined in Refs.~\cite{Cerf,Horod}), since they will be used repeatedly in what follows. 
Since the Universal Inverter is a positive 
operator which preserves Hermiticity, so is $\Lambda_d$. Nevertheless, 
$\Lambda_d$ is {\it not} completely positive. In fact, one can prove that it can be decomposed as a completely positive map $\Lambda_d^{CP}$ supplemented with the transpose -- or time reversal -- map $T$, i.e., $\Lambda_d = \Lambda_d^{CP}T$  \cite{Horod}. Its action on an arbitrary operator $\sigma$ is given by
\begin{equation}\label{defconc}
\Lambda_d (\sigma)=\mathrm{Tr}(\sigma)I\otimes I-\sigma_A\otimes I - I\otimes \sigma_B + \sigma
\end{equation}
where $\sigma_A=\mathrm{Tr}_B (\sigma)$ and $\sigma_B=\mathrm{Tr}_A(\sigma)$ are the reduction of the operator $\sigma$ on $A$ and $B$ respectively \cite{Cerf,Horod}. The map $\Lambda_d$ is trace-increasing as
\begin{equation}\label{traceinc}
\mathrm{Tr}(\Lambda_d(\sigma))=(d-1)^2 \; \mathrm{Tr}(\sigma)
\end{equation}
Finally, one can easily check that expressions such as $\mathrm{Tr}(\rho\Lambda_d(\sigma))$ are symmetric with respect to the
interchange between $\sigma$ and $\rho$, that is,
\begin{equation}\label{sym}
\mathrm{Tr}(\rho\Lambda_d(\sigma))=\mathrm{Tr}(\sigma\Lambda_d(\rho))
\end{equation}

Now, coming back to the generalized concurrence in dimension $d$, we note first
that by expressing relation (\ref{defconc}) in the case of a pure state $\sigma=\vert\Psi\rangle\langle\Psi\vert$, we can rewrite the concurrence as \begin{equation}
C(\Psi)=\sqrt{2(1-\mathrm{Tr}(\rho^2_A))}.
\end{equation} 
Hence, the concurrence ranges from $0$ for separable states 
to $\sqrt{2(d-1)/d}$ for maximally entangled ones. 
Next, to address the problem of the concurrence of superpositions, we can plug 
Eq.~(\ref{sup}) into the definition (\ref{conc}) of the concurrence, 
or actually the square of the concurrence (to get rid of the unnecessary square root). We can develop this expression, resulting in a simple yet lengthy summation over 16 different terms. Out of these 16 terms, many are identical as can be proven using relation (\ref{sym}). For example, 
\begin{align}
\nonumber\langle\phi\vert\Lambda_d(\vert\phi\rangle\langle\varphi\vert)\vert\phi\rangle
&= \mathrm{Tr}(\vert\phi\rangle\langle\phi\vert \Lambda_d(\vert\phi\rangle\langle\varphi\vert) )\\
\nonumber &= \mathrm{Tr}(\vert\phi\rangle\langle\varphi\vert \Lambda_d(\vert\phi\rangle\langle\phi\vert)  )\\
&= \langle\varphi\vert\Lambda_d(\vert\phi\rangle\langle\phi\vert)\vert\phi\rangle
\end{align}
In addition, one can also show that $\langle\phi\vert\Lambda_d(\vert\phi\rangle\langle\varphi\vert)\vert\varphi\rangle= \langle\varphi\vert\Lambda_d(\vert\phi\rangle\langle\phi\vert)\vert\varphi\rangle$ holds, as well as its counterpart  $\langle\varphi\vert\Lambda_d(\vert\varphi\rangle\langle\phi\vert)\vert\phi\rangle= \langle\phi\vert\Lambda_d(\vert\varphi\rangle\langle\varphi\vert)\vert\phi\rangle$. To do so, first develop both sides using the expression (\ref{defconc}) for $\Lambda_d$, write $\vert\phi\rangle$ and $\vert\varphi\rangle$ in a given basis, say the Schmidt basis of $\vert\phi\rangle$ for simplicity, and then prove that all four terms are identical. We finally obtain a general expression for the square of the concurrence of a superposition state, 
\begin{align} \label{cc}
\nonumber C^2&(\Psi)=\vert\alpha\vert^4 C^2(\phi)+\vert\beta\vert^4 C^2(\varphi)+ 4\vert\alpha\beta\vert^2\langle\varphi\vert\Lambda_d(\vert\phi\rangle\langle\phi\vert)\vert\varphi\rangle\\
\nonumber &+ 2\vert\alpha\vert^2\big(\alpha^*\beta\langle\phi\vert\Lambda_d(\vert\phi\rangle\langle\phi\vert)\vert\varphi\rangle + \alpha\beta^*\langle\varphi\vert\Lambda_d(\vert\phi\rangle\langle\phi\vert)\vert\phi\rangle\big)\\
\nonumber &+ 2\vert\beta\vert^2\big(\alpha^*\beta\langle\phi\vert\Lambda_d(\vert\varphi\rangle\langle\varphi\vert)\vert\varphi\rangle + \alpha\beta^*\langle\varphi\vert\Lambda_d(\vert\varphi\rangle\langle\varphi\vert)\vert\phi\rangle\big)\\
&+ \big((\alpha^*\beta)^2\langle\phi\vert\Lambda_d(\vert\varphi\rangle\langle\phi\vert)\vert\varphi\rangle + (\alpha\beta^*)^2\langle\varphi\vert\Lambda_d(\vert\phi\rangle\langle\varphi\vert)\vert\phi\rangle\big)
\end{align}

\subsection{Biorthogonal states} 

When the two component states $\vert\phi\rangle$ and $\vert\varphi\rangle$ 
are biorthogonal, that is, when 
\begin{align}\label{biorth}
\nonumber \mathrm{Tr}_A(\mathrm{Tr}_B(\vert\phi\rangle\langle\phi\vert)
\mathrm{Tr}_B(\vert\varphi\rangle\langle\varphi\vert))&=0\\
\mathrm{Tr}_B(\mathrm{Tr}_A(\vert\phi\rangle\langle\phi\vert)\mathrm{Tr}_A(\vert\varphi\rangle\langle\varphi\vert))&=0
\end{align}
an exact expression for the entanglement of a superposition can be derived based on the entanglement of formation, see Ref.~\cite{Linden}. As stated in the following theorem, one can also find an exact value for the concurrence of the superposition. 
\\
\indent\textbf{Theorem 1: Biorthogonal states.} If $\vert\phi\rangle$ and $\vert\varphi\rangle$ satisfy conditions (\ref{biorth}) and if $\vert\alpha\vert^2+\vert\beta\vert^2=1$, then the concurrence of the superposition $\vert\Psi\rangle=\alpha\vert\phi\rangle+\beta\vert\varphi\rangle$ is given by
\begin{align}
C(\Psi)=\sqrt{\vert\alpha\vert^4 C^2(\phi)+\vert\beta\vert^4 C^2(\varphi)
+4\vert\alpha\beta\vert^2}
\end{align}
\textit{Proof:} In addition to conditions (\ref{biorth}), we also note that the reductions on $A$ and $B$ of the operators $\vert\phi\rangle\langle\varphi\vert$ and  $\vert\varphi\rangle\langle\phi\vert$ are equal to zero. Hence, one can easily check, using relation (\ref{defconc}), that all the terms of (\ref{cc}) are zero except from the first three. The equality follows from the fact that $\langle\varphi\vert\Lambda_d(\vert\phi\rangle\langle\phi\vert)\vert\varphi\rangle=1$ when the states are biorthogonal [as can be checked by Eq.~(\ref{defconc}) again]. $\square$\\

\subsection{Orthogonal states} 

Let us now consider the less restrictive case of $\vert\phi\rangle$ and $\vert\varphi\rangle$ being orthogonal but not necessarily biorthogonal. As we will prove in the next theorem, one can derive a simple upper bound which nicely generalizes the upper bound (\ref{sup_2}) that we had found for qubits.\\
\indent\textbf{Theorem 2: Upper bound.} Let Alice and Bob each have a qudit, and let $\vert\phi\rangle$ and $\vert\varphi\rangle$ be orthogonal states, the concurrence of the superposition $\vert\Psi\rangle=\alpha\vert\phi\rangle+\beta\vert\varphi\rangle$ with $\vert\alpha\vert^2+\vert\beta\vert^2=1$ satisfies
\begin{align} \label{upd}
C(\Psi)\leq\vert\alpha\vert^2 C(\phi) + \vert\beta\vert^2 C(\varphi)+ 2\vert\alpha\beta\vert
\end{align}
\textit{Proof:} First, remember that the map $\Lambda_d$, while it is
not completely positive, can be written as the transpose map $T$ followed by a completely positive map $\Lambda^\mathrm{CP}_d$, 
i.e. $\Lambda_d=\Lambda^\mathrm{CP}_d T$. Since $\Lambda^\mathrm{CP}_d$
is completely positive, it has an operator-sum representation based on a set of operators $\{A_k\}$. Hence, for an arbitrary operator $\sigma$, the action of $\Lambda_d$ can be written as
\begin{equation}\label{opsum}
\Lambda_d (\sigma) = \Lambda^\mathrm{CP}_d (T(\sigma))
=\sum_k A_k \sigma^T A_k^\dagger
\end{equation}
where the operators $A_k$ satisfy $\sum_k A_k A_k^\dagger=(d-1)^2 I$ according to (\ref{traceinc}). We can now make use of this decomposition to bound the different terms of (\ref{cc}). Considering $\langle\phi\vert\Lambda_d(\vert\varphi\rangle\langle\phi\vert)\vert\varphi\rangle$, for example, one can prove that
\begin{align}
\vert\langle\phi\vert\Lambda_d&(\vert\varphi\rangle\langle\phi\vert)\vert\varphi\rangle\vert = \big\vert \sum_k\langle\phi\vert A_k \vert \phi^*\rangle\langle\varphi^*\vert A_k^\dagger \vert
 \varphi\rangle\big\vert\\
\nonumber &\leq \sum_k\vert\langle\phi\vert A_k \vert \phi^*\rangle\vert\vert\langle\varphi^*\vert A_k^\dagger \vert
 \varphi\rangle\vert\\
\nonumber &\leq \sqrt{\sum_k\vert\langle\phi\vert A_k \vert \phi^*\rangle\vert^2}\sqrt{\sum_k\vert\langle\varphi^*\vert A_k^\dagger \vert
 \varphi\rangle\vert^2}\\
\nonumber &= \sqrt{\langle\phi\vert\Lambda_d(\vert\phi\rangle\langle\phi\vert)\vert\phi\rangle} \sqrt{\langle\varphi\vert\Lambda_d(\vert\varphi\rangle\langle\varphi\vert)\vert\varphi\rangle}\\
\nonumber &=C(\phi)C(\varphi)
\end{align}
where in the first line we have made use of relation (\ref{opsum}) and expressed the transposition in a fixed basis of the operator $\vert\varphi\rangle\langle\phi\vert$ as $(\vert\varphi\rangle\langle\phi\vert)^T=\vert\phi^*\rangle\langle\varphi^*\vert$. The second and third lines follow respectively from the triangle inequality and Holder's inequality (see e.g.\cite{holder}). The last two equalities result from Eq.~(\ref{opsum}) again and the definition of the concurrence of $\vert\phi\rangle$ and $\vert\varphi\rangle$. We can repeatedly use the same argument on the other terms and prove
\begin{align} \label{boundnorm}
\nonumber \vert\langle\varphi\vert\Lambda_d(\vert\phi\rangle\langle\varphi\vert)\vert\phi\rangle\vert &\leq C(\phi)C(\varphi)\\
\nonumber \vert\langle\phi\vert\Lambda_d(\vert\phi\rangle\langle\phi\vert)\vert\varphi\rangle\vert&\leq C(\phi)\sqrt{\langle\varphi\vert\Lambda_d(\vert\phi\rangle\langle\phi\vert)\vert\varphi\rangle}\\
\nonumber\vert\langle\varphi\vert\Lambda_d(\vert\phi\rangle\langle\phi\vert)\vert\phi\rangle\vert&\leq C(\phi)\sqrt{\langle\varphi\vert\Lambda_d(\vert\phi\rangle\langle\phi\vert)\vert\varphi\rangle}\\
\nonumber\vert\langle\varphi\vert\Lambda_d(\vert\varphi\rangle\langle\varphi\vert)\vert\phi\rangle\vert&\leq C(\varphi)\sqrt{\langle\varphi\vert\Lambda_d(\vert\phi\rangle\langle\phi\vert)\vert\varphi\rangle}\\
\vert\langle\phi\vert\Lambda_d(\vert\varphi\rangle\langle\varphi\vert)\vert\varphi\rangle\vert&\leq C(\varphi)\sqrt{\langle\varphi\vert\Lambda_d(\vert\phi\rangle\langle\phi\vert)\vert\varphi\rangle}
\end{align}
Finally, all we need to do is to bound $C^2(\Psi)$ by the sum of the modulus of each of the terms of (\ref{cc}), then bound each term individually by use of the previous relations. This gives
\begin{align}\label{updsquare}
C^2(\Psi)\leq \Big( \vert\alpha\vert^2 C(\phi) 
&+\vert\beta\vert^2 C(\varphi) \nonumber \\
&+ 2\vert\alpha\beta\vert \sqrt{\langle\varphi\vert\Lambda_d(\vert\phi\rangle\langle\phi\vert)\vert\varphi\rangle} \Big)^2
\end{align}
The conclusion follows from the value of $\langle\varphi\vert\Lambda_d(\vert\phi\rangle\langle\phi\vert)\vert\varphi\rangle$, which can be bounded using (\ref{defconc}), that is,
\begin{align} \label{sqrt_orth}
\nonumber \langle\varphi\vert\Lambda_d(\vert\phi\rangle\langle\phi\vert)\vert\varphi\rangle &= 1-\mathrm{Tr}_A(\rho^A_\phi\rho^A_\varphi)-\mathrm{Tr}_B(\rho^B_\phi\rho^B_\varphi)\\
&\leq 1
\end{align}
where $\rho^{A,B}_\phi=\mathrm{Tr}_{B,A}(\vert\phi\rangle\langle\phi\vert)$ and $\rho^{A,B}_\varphi=\mathrm{Tr}_{B,A}(\vert\varphi\rangle\langle\varphi\vert)$. $\square$\\

Let us make a few comments at this point. First, it is interesting to see how the structure of our upper bound closely resemble the one derived for the entanglement of formation in \cite{Linden}. Both bounds are the sum of two terms: the first one is the average of the entanglement of $\vert\phi\rangle$ and $\vert\varphi\rangle$ (up to a factor), and the second one is a function of $\alpha$ and $\beta$ only, which takes its maximum value when $\alpha=\beta$.

Another comment we should make is related to the definition of the I-concurrence. Although the upper bound (\ref{upd}) resembles much the bound that we have derived for qubits, Eq.~(\ref{sup_2}), the third term lacks the correction factor in $\delta$ that we had deduced from (\ref{relorth}). To understand this difference, remember that the Universal Inverter is defined up to a scaling factor $\nu_d$. 
As noted in Ref.~\cite{Rungta}, the definition of a generalized measure of entanglement such as the concurrence requires $\nu_d$ to be independent of the dimension, otherwise the concurrence could be changed simply by adding an extra dimension that is not used to one of the subsystems. Unfortunately, this leads to a Universal Inverter which is not trace-preserving, and with the choice $\nu_d=1$, the condition (\ref{relorth}) becomes
\begin{equation}
\sqrt{\langle\varphi\vert\Lambda_d(\vert\phi\rangle\langle\phi\vert)\vert\varphi\rangle}\leq \sqrt{(d-1)^2-\delta^2}
\end{equation}
This bound becomes useless whenever $d>2$, which explains the discrepancy between (\ref{sup_2}) and (\ref{upd}).
If, on the other hand, we had chosen $\nu_d=1/(d-1)$ to make the Universal Inverter trace-preserving (at the expense of an ill-defined I-concurrence), relation (\ref{relorth}) would still be valid for $d>2$ and our upper bound (\ref{sup_2}) would hold regardless of the dimension. Thus, with respect to our chosen definition of the I-concurrence, our upper bound for qudits is consistent with the bound we had previously derived for qubits. \\

Let us now derive a lower bound for the concurrence of the superposition
of orthogonal states, generalizing the bound obtained for qubits, Eq.~(\ref{inf_2}). To do so, a natural approach would be to start again from Eq.~(\ref{cc}), apply the inverse triangle inequality to separate between the first three terms, which are always positive, and the rest of the expression, which can take negative values. This negative part could then be upper bounded by repeated uses of the triangle inequality, followed by the introduction of relations (\ref{boundnorm}) to bound the norm of each term separately. Unfortunately, this approach would lead to a rather bad and complicated bound. However, as stated in the next theorem, one can nevertheless find a simple yet meaningful bound by use of different approach.\\

\indent\textbf{Theorem 3: Lower bound.} Let Alice and Bob each have a qudit, and let $\vert\phi\rangle$ and $\vert\varphi\rangle$ be orthogonal, the concurrence of the superposition $\vert\Psi\rangle=\alpha\vert\phi\rangle+\beta\vert\varphi\rangle$, with $\vert\alpha\vert^2+\vert\beta\vert^2=1$ satisfies
\begin{align} \label{lod}
C(\Psi)\geq\big\vert\vert\alpha\vert^2 C(\phi) - \vert\beta\vert^2 C(\varphi)\big\vert - 2\vert\alpha\beta\; \vert(1+\delta)
\end{align}
where $\delta=\min\big(\vert\frac{\beta}{\alpha}\, \vert C(\varphi),\vert\frac{\alpha}{\beta}\, \vert C(\phi)\big)$.\\

\textit{Proof:} First, we note that Eq.~(\ref{sup}) can be rewritten as
\begin{align} \label{rewritten}
\vert\phi\rangle=\frac{\alpha^*}{\vert\alpha\vert^2}\vert\Psi\rangle - \frac{\alpha^2\beta}{\vert\alpha\vert^2}\vert\varphi\rangle
\end{align}
Next, we make use of the fact that the bound (\ref{updsquare}) was derived based solely on the properties of the map $\Lambda_d$ and on the normalization condition $\Vert \Psi \Vert = 1$.  We did not make explicit use of the orthogonality of  $\vert\phi\rangle$ and $\vert\varphi\rangle$, neither did we use the condition $\vert\alpha\vert^2+\vert\beta\vert^2=1$ in the derivation. It follows that the bound (\ref{updsquare}) can be applied to Eq.~(\ref{rewritten}),
that is,
\begin{align*}
C(\phi)&\leq\frac{1}{\vert\alpha\vert^2}C(\Psi) + \vert\beta\vert^2 C(\varphi) + 2\frac{\vert\beta\vert}{\vert\alpha\vert^2}\sqrt{\langle\Psi\vert\Lambda_d(\vert\varphi\rangle\langle\varphi\vert)\vert\Psi\rangle}
\end{align*}
from which we deduce
\begin{align} \label{inv_lb}
C(\Psi)&\geq \vert\alpha\vert^2 C(\phi) - \vert\beta\vert^2 C(\varphi) - 2\vert\beta\vert\sqrt{\langle\Psi\vert\Lambda_d(\vert\varphi\rangle\langle\varphi\vert)\vert\Psi\rangle}
\end{align}
Furthermore, we can plug Eq.~(\ref{sup}) into $\langle\Psi\vert\Lambda_d(\vert\varphi\rangle\langle\varphi\vert)\vert\Psi\rangle $  in order to bound the last term of the right-hand side
of Eq.~(\ref{inv_lb}), that is,
\begin{align}
\nonumber \langle\Psi\vert\Lambda_d(\vert\varphi\rangle\langle\varphi\vert)\vert\Psi\rangle &=\vert\alpha\vert^2 \langle\phi\vert\Lambda_d(\vert\varphi\rangle\langle\varphi\vert)\vert\phi\rangle\\
\nonumber &+ \vert\beta\vert^2 \langle\varphi\vert\Lambda_d(\vert\varphi\rangle\langle\varphi\vert)\vert\varphi\rangle \\
\nonumber &+ 2\Re e(\vert\alpha\beta^*\vert\langle\varphi\vert\Lambda_d(\vert\varphi\rangle\langle\varphi\vert)\vert\phi\rangle )\\
\nonumber &\leq \big(\vert\alpha\vert\sqrt{\langle\phi\vert\Lambda_d(\vert\varphi\rangle\langle\varphi\vert)\vert\phi\rangle} + \vert\beta\vert C(\varphi) \big)^2\\
&\leq \big(\vert\alpha\vert + \vert\beta\vert C(\varphi) \big)^2
\end{align}
where we have bounded the real part of the complex number by its modulus and used the third relation of (\ref{boundnorm}) to obtain the first inequality. 
The last line follows directly from Eq.~(\ref{sqrt_orth}). Thus, we get
\begin{align} \label{bound-1}
C(\Psi)\geq \vert\alpha\vert^2 C(\phi) - \vert\beta\vert^2 C(\varphi) - 2\vert\alpha\beta\vert (1+\vert \frac{\beta}{\alpha}\vert C(\varphi))
\end{align}
Alternatively, we can express $\vert\varphi\rangle$ as a superposition of $\vert\Psi\rangle$ and $\vert\phi\rangle$ instead of starting from Eq.~(\ref{rewritten}), which would result into the dual relation
\begin{align} \label{bound-2}
C(\Psi)\geq \vert\beta\vert^2 C(\varphi) - \vert\alpha\vert^2 C(\phi) - 2\vert\alpha\beta\vert (1+\vert \frac{\alpha}{\beta}\vert C(\phi))
\end{align}
Combining these two bounds (\ref{bound-1}) and (\ref{bound-2}) and noticing that $\vert\alpha\vert^2 C(\phi) - \vert\beta\vert^2 C(\varphi)>0$ is equivalent to $\vert\frac{\beta}{\alpha}\vert  C(\varphi)<\vert\frac{\alpha}{\beta}\vert C(\phi)$ leads to the conclusion. $\square$\\

We do not know whether Eq.~(\ref{lod}) is the best possible lower
bound, and suspect that a more appropriate expression, if it exists, would probably have the correction factor $\delta$ equal to zero as it would then generalize the bound obtained for qubits. Nevertheless, our bound has the desired shape and is close to this guessed optimal when $\delta$ is small, that is, when the concurrence of one of the superposed state is small, or when the superposition is strongly unbalanced, i.e. $\vert\alpha\vert\gg\vert\beta\vert$ or $\vert\beta\vert\gg\vert\alpha\vert$.  As a last comment, let us note that this lower bound is useful (i.e., provides a value above zero) whenever 
\begin{align}
C(\phi)&>3\vert\frac{\beta}{\alpha}\vert^2 C(\varphi) + 2\vert\frac{\beta}{\alpha}\vert \quad \text{or}\\
C(\varphi)&>3\vert\frac{\alpha}{\beta}\vert^2 C(\phi) + 2\vert\frac{\alpha}{\beta}\vert
\end{align}

\begin{figure}[t]
\centering
\includegraphics[scale=0.5]{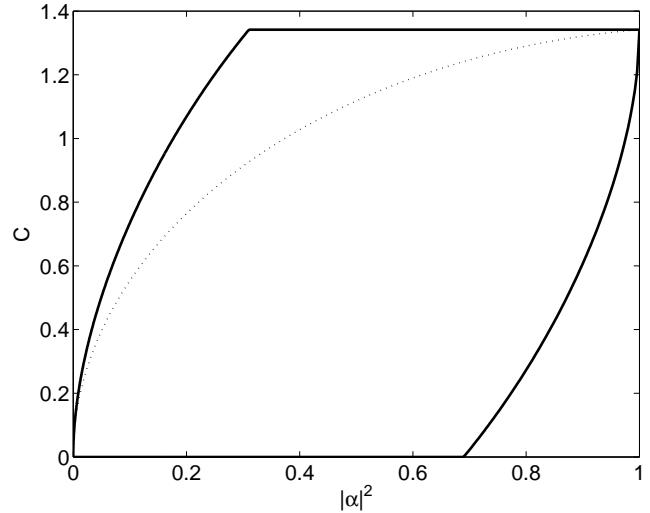}
\caption{Concurrence of $\vert\Psi\rangle=\alpha\vert\phi\rangle+\beta\vert\varphi\rangle$,
with $\vert\phi\rangle$ and $\vert\varphi\rangle$ 
as defined in the text. The dotted line is the exact value of $C(\Psi)$, while the two solid lines correspond to the upper and lower bounds derived from relation (\ref{upd}) and (\ref{lod}).}
\label{fig:C}
\end{figure}

In Fig.~\ref{fig:C}, we illustrate the upper and lower bounds for the
case of two orthogonal component states in dimension $d=10$, namely
\begin{align}
\vert\phi\rangle &= \frac{1}{10} \sum_{i,j=0}^9 |i,j\rangle \nonumber\\ 
\vert\varphi\rangle &= \frac{1}{\sqrt{10}} \sum_{i=0}^9 |i,i\rangle  
\end{align}
In general, we notice that the bounds in high dimensions are less constraining
than for $d=2$. This is natural as for a larger dimension, the set of
states with a given value of the concurrence gets larger, hence the range
of possible values for the concurrence of the superposition of two states with a fixed concurrence becomes wider.

\subsection{Arbitrary states} 

Let us show finally that the previous theorems can be nicely generalized to the completely general situation of the component states $\vert\phi\rangle$ and $\vert\varphi\rangle$ being arbitrary.\\

\indent\textbf{Theorem 4: Arbitrary states.} Let $\vert\phi\rangle$ and $\vert\varphi\rangle$ be normalized but arbitrary, and let $\vert\alpha\vert^2+\vert\beta\vert^2=1$, the concurrence of $\vert\Psi'\rangle$, the normalized version of the superposition $\vert\Psi\rangle=\alpha\vert\phi\rangle+\beta\vert\varphi\rangle$, satisfies
\begin{align*}
\Vert \Psi\Vert^2 C(\Psi')&\leq\vert\alpha\vert^2 C(\phi) + \vert\beta\vert^2 C(\varphi)+ 2\vert\alpha\beta\vert\sqrt{1+\vert\langle\phi\vert\varphi\rangle\vert^2}\\
\nonumber \Vert \Psi\Vert^2 C(\Psi')&\geq\big\vert\vert\alpha\vert^2 C(\phi) - \vert\beta\vert^2 C(\varphi)\big\vert\\
 &- 2\vert\alpha\beta\vert(\sqrt{1+\vert\langle\phi\vert\varphi\rangle\vert^2}+\delta)
\end{align*}
where $\delta=\min\big(\vert\frac{\beta}{\alpha}\vert C(\varphi),\vert\frac{\alpha}{\beta}\vert C(\phi)\big)$.\\

\textit{Proof:} First, we note that $\Vert\Psi\Vert^4 C^2(\Psi')=\langle\Psi\vert\Lambda_d(\vert\Psi\rangle\langle\Psi\vert)\vert\Psi\rangle$. Next, we remember that the derivation of Theorems 2 and 3 was mostly based on the property of the map $\Lambda_d$ only, for a normalized state $\vert\Psi\rangle$. In particular, Eqs.~(\ref{updsquare}) and (\ref{inv_lb}) as well as the dual of the latter equation were derived without explicitely assuming  $\vert\phi\rangle$ and $\vert\varphi\rangle$ to be orthogonal. One can thus easily check that these inequalities hold provided that $C(\Psi)$ is replaced by $\Vert\Psi\Vert^2 C(\Psi')$. Furthermore, when $\vert\langle\phi\vert\varphi\rangle\vert\neq0 $, Eq.~(\ref{sqrt_orth}) should be replaced by
\begin{align}
\langle\varphi\vert\Lambda_d(\vert\phi\rangle\langle\phi\vert)\vert\varphi\rangle \leq 1+\vert\langle\phi\vert\varphi\rangle\vert^2
\end{align}
as can be deduced from (\ref{defconc}). Combining these relations directly leads to the conclusion. $\square$\\

\section{Conclusion}

We have investigated the concurrence of the quantum superposition of two bipartite pure states in arbitrary dimension. The concurrence being 
an entanglement measure that is distinct from the entanglement of formation
in dimensions higher than 2, our study complements that of Ref.~\cite{Linden}.
We have derived simple relations between the concurrence of the superposition state and the concurrence of its two component states. When the scalar product of these two states is known, our method provides both a lower and an upper bound on the concurrence of the superposition state.  These bounds take particularly simple forms when this scalar product
is zero, i.e., when the two component states are orthogonal.
In dimension 2, we have checked that our upper bound is typically tighter 
than the bound of Ref.~\cite{Linden}. Finally,
in the special case where the component states are biorthogonal, we have 
derived an exact expression for the concurrence of the superposition. 
\\

\section{Acknowledgments}
J.N. acknowledges support from the Belgian
FRIA foundation. We also acknowledge financial support from the IUAP programme of the 
Belgian government under project Photonics@be and from the EU under projects 
QAP. \\

\textit{Note:} While completing this work, a related paper on the
entanglement of quantum superpositions has appeared, where the generalization
to a multipartite setting and an arbitrary entanglement witness is considered
\cite{cavalcanti}. Some of the results obtained there are in perfect
agreement with ours.

\end{document}